\documentclass[preprint2]{aastex}
\usepackage{rotating}
\usepackage{graphicx}
\usepackage{grffile}
\usepackage{multicol}
\usepackage{longtable}

\begin{document}
\setcounter{secnumdepth}{2}
\title{The TRENDS High-Contrast Imaging Survey. III. \\ A Faint White Dwarf Companion Orbiting HD 114174}
\author{Justin R. Crepp\altaffilmark{1}, John Asher Johnson\altaffilmark{2}, Andrew W. Howard\altaffilmark{3}, Geoffrey W. Marcy\altaffilmark{5}, Alexandros Gianninas\altaffilmark{4}, Mukremin Kilic\altaffilmark{4}, Jason T. Wright\altaffilmark{6,7}}
\email{jcrepp@nd.edu} 
\altaffiltext{1}{Department of Physics, University of Notre Dame, 225 Nieuwland Science Hall, Notre Dame, IN, 46556, USA}
\altaffiltext{2}{Department of Planetary Science, California Institute of Technology, 1200 E. California Blvd., Pasadena, CA 91125, USA}
\altaffiltext{3}{Institute for Astronomy, University of Hawaii, 2680 Woodlawn Drive, Honolulu, HI 96822}
\altaffiltext{4}{Department of Physics and Astronomy, University of Oklahoma, Norman, OK 73019}
\altaffiltext{5}{Department of Astronomy, University of California, Berkeley, CA 94720, USA} 
\altaffiltext{6}{Department of Astronomy \& Astrophysics, The Pennsylvania State University, University Park, PA 16802, USA} 
\altaffiltext{7}{Center for Exoplanets and Habitable Worlds, The Pennsylvania State University, University Park, PA 16802, USA}

\begin{abstract} 
The nearby Sun-like star HD~114174 exhibits a strong and persistent Doppler acceleration indicating the presence of an unseen distant companion. We have acquired high-contrast imaging observations of this star using NIRC2 at Keck and report the direct detection of the body responsible for causing the ``trend". HD~114174~B has a projected separation of $692\pm9$ mas ($18.1$ AU) and is $10.75\pm0.12$ magnitudes (contrast of $5\times10^{-5}$) fainter than its host in the $K$-band, requiring aggressive point-spread function subtraction to identify. Our astrometric time baseline of 1.4 years demonstrates physical association through common proper motion. We find that the companion has absolute magnitude, $M_J=13.97\pm0.11$, and colors, $J-K= 0.12\pm0.16$ mag. These characteristics are consistent with an $\approx$T3 dwarf, initially leading us to believe that HD~114174~B was a substellar object. However, a dynamical analysis that combines radial velocity measurements with available imaging data indicates a minimum mass of $0.260\pm0.010M_{\odot}$. We conclude that HD~114174~B must be a white dwarf. Assuming a hydrogen-rich composition, atmospheric and evolutionary model fits yield an effective temperature $T_{\rm eff} = 8200\pm4000$ K, surface gravity $\log g=8.90\pm0.02$, and cooling age of $t_c\approx3.4$ Gyr, which is consistent with the $4.7^{+2.3}_{-2.6}$ Gyr host star isochronal age estimate. HD~114174~B is a benchmark object located only $26.14\pm0.37$ pc from the Sun. It may be studied at a level of detail comparable to Sirius and Procyon, and used to understand the link between the mass of white dwarf remnants with that of their progenitors. 
\end{abstract}

\section{INTRODUCTION}\label{sec:intro}
The TRENDS ({\bf T}a{\bf R}getting b{\bf EN}chmark-objects with {\bf D}oppler {\bf S}pectroscopy) high-contrast imaging program uses precise radial velocity (RV) measurements obtained over a long time baseline to identify promising targets for follow-up observations using adaptive optics (AO) (Crepp et al. 2012b, Crepp et al. 2012c). In addition to diffraction-limited imaging, TRENDS also uses coronagraphy, point-spread function (PSF) subtraction, and $\approx$1 hour integration times to achieve deep contrast at near-infrared wavelengths \citep{marois_06,lafreniere_07,crepp_10}. The primary goal of TRENDS is to image and characterize benchmark M-dwarfs, brown dwarfs, and (ultimately) extrasolar planets, using a combination of complementary substellar companion detection techniques, in order to calibrate theoretical evolutionary models and theoretical spectral models of cool dwarf atmospheres.

Precise Doppler measurements for TRENDS are obtained at Keck Observatory using the HIgh Resolution Echelle Spectrometer (HIRES; \citet{vogt_94}) as part of the California Planet Search program \citep{howard_10}. Targets with the longest RV time baselines have complementary data taken at Lick Observatory using the 0.6m Coude Auxiliary Telescope (CAT) and Shane 3m Telescope with the Hamilton spectrometer \citep{vogt_87}. High-contrast imaging follow-up observations are conducted at Keck using NIRC2 (instrument PI: Keith Matthews) with the Keck II AO system \citep{wizinowich_00}, and at Palomar using Project 1640 with the PALM-3000 high-order AO system \citep{hinkley_11_PASP} . 

\citealt{crepp_12b} (hereafter TRENDS I) reported the discovery of three benchmark high-mass ratio binaries, HD~53665, HD~68017, and HD~71881, each companion having an age and metallicity estimate inferred from the solar-type primary \citep{crepp_12b}. Arguably the most interesting object of the three, HD~68017~B, has a RV time baseline of 15 years and shows significant orbit curvature. With a projected separation of only 13.0 AU, it will be possible to measure a precise dynamical mass in the next several years. The model-dependent mass of HD~68017~B from relative photometry is $m_{\rm model}=168\pm21M_J$ ($0.16\pm0.02M_{\odot}$). 

\citealt{crepp_13a} (TRENDS II) reported the direct detection of an $m_{\rm model}=0.52\pm0.04M_{\odot}$ tertiary companion orbiting the 84 day period, single-line spectroscopic binary HD~8375~AB. With a separation of $0\farcs3$, HD~8375~C has a brightness comparable to residual scattered starlight from the inner binary pair prior to PSF subtraction. The HD~8375 system represents one of many hierarchical triples in the solar neighborhood that are known to exist, through statistical analysis of multiplicity fractions for stellar systems within $d<10$ pc, but so often evade direct detection \citep{tokovinin_04}.

In this paper (TRENDS III), we report the discovery of a faint object orbiting the nearby, $d=26.14\pm0.37$ pc (\citealt{van_leeuwen_07}), G5 star HD~114174 (LHS~2681, HIP~64150, GJ~9429). The companion, HD~114174~B, cannot be seen without significant on-source integration time and PSF subtraction. It has a projected separation ($0.69\arcsec=18.1$ AU) comparable to the semimajor axis ($a=18.3$ AU) of the brown dwarf HR~7672~B, which recently yielded a dynamical mass with a fractional error of only 4\% \citep{crepp_12a}. Initially characterized as a substellar object based on its brightness and colors, we find that HD~114174~B must instead be a white dwarf, in order to establish a self-consistent interpretation with long-term Doppler RV measurements. 

Few ``Sirius-like" systems, i.e., directly imaged white dwarfs around solar-type stars, are presently known -- approximately one dozen within 20 pc \citep{holberg_09}. Our serendipitous discovery of a compact object orbiting HD~114174 thus constitutes a new member of a small collection of nearby, directly detected benchmark white dwarfs for which we can determine physical properties independent of spectro-photometric measurements. Forthcoming measurements will ultimately lead to tight dynamical constraints on its orbit and mass. More than 16 years of RV measurements are already in hand. 

\begin{figure*}[!t]
\begin{center}
\includegraphics[height=5in]{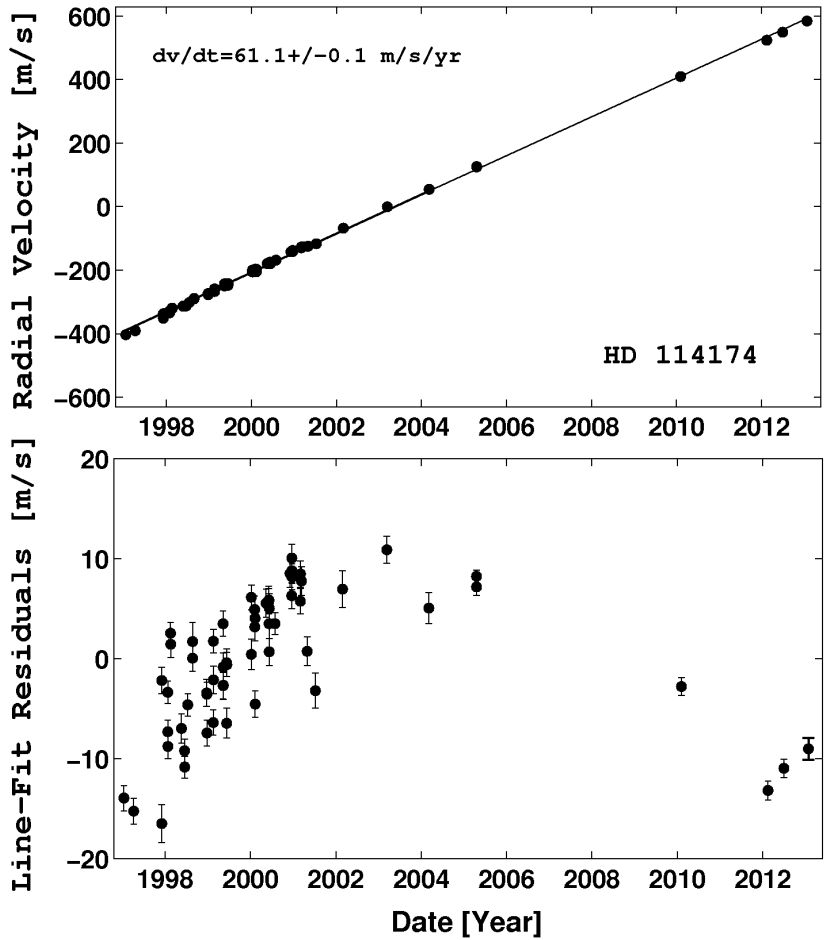} 
\caption{Relative radial velocity measurements of HD~114174 from the years 1997 through 2013. (Top) A long-term Doppler acceleration of $61.1\pm0.1$ m/s/yr implies the presence of a companion with a long period. (Bottom) A linear fit model results in large residuals indicating systemic curvature (change in the acceleration). Thus, it will soon be possible to calculate the three-dimensional orbit and dynamical mass of HD~114174~B.} 
\end{center}\label{fig:rvs}
\end{figure*} 

\section{OBSERVATIONS}

\subsection{High-Resolution Stellar Spectroscopy}

\subsubsection{Doppler Measurements}
We obtained RV data of HD~114174 using HIRES at Keck. The standard iodine cell referencing method was used to calibrate instrument drifts and measure precise Doppler shifts \citep{marcy_butler_92}. Our Doppler observations began on January 14, 1997 UT. Several years of measurements revealed that the star exhibits a (mostly) linear acceleration indicative of a distant companion. Fig. 1 shows Doppler observations taken over an 16.0 year timeframe. RV values, epochs, and measurement uncertainties are listed in Table 1. 

Treating the Doppler trend as purely linear and using a Markov Chain Monte Carlo (MCMC) fitting procedure, we find an acceleration of $61.1\pm0.1\:\rm{m\:s}^{-1}\rm{yr}^{-1}$. This analysis includes a differential offset between measurements obtained before and after the summer 2004 HIRES detector upgrade, which we treat as a free parameter. Closer inspection of the data shows that the RV slope is slowly changing with time. For instance, the bottom panel of Fig. 1 displays RV residuals once the measurements are subtracted from a linear model. Continued Doppler monitoring will allow us to place tight constraints on the companion orbit and dynamical mass, as soon as the direct imaging astrometry shows curvature. By performing a joint RV-astrometry analysis using presently available data, we derive a lower-limit for the mass of HD~114174~B (Section 5.1) and determine a range of plausible orbital periods (Section 5.4). 

\subsubsection{Stellar Properties}
Stellar template spectra, taken with the iodine gas cell removed from the optical path, were analyzed using the LTE spectral synthesis code {\it Spectroscopy Made Easy} (SME) described in \citet{valenti_96} and \citet{valenti_fischer_05}. SME provides an estimate of the stellar effective temperature ($T_{\rm eff}$), surface gravity ($\log g$), metallicity ($\mbox{[Fe/H]}$), and projected rotational velocity ($v \sin i$). The estimated physical properties of HD~114174 derived from spectral fitting are shown in Table 3. 

HD~114174 is listed in the SIMBAD database as a G5(IV) subgiant. However, its location with respect to the mean Hipparcos main-sequence \citep{wright_05} and spectroscopic $\log g=4.51\pm0.06$ value indicate the star is a main-sequence dwarf. Placing HD~114174 on a Hertzsprung-Russell diagram, we estimate a mass of $1.05\pm0.05M_{\odot}$ and age of $4.7^{+2.3}_{-2.6}$ Gyr by comparing spectroscopic parameters to Yonsei-Yale theoretical isochrone tracks. Despite its close distance, HD~114174 would rarely, if ever, make typical high-contrast imaging target lists, which nominally only select stars younger than $\approx$1 Gyr \citep{crepp_johnson_11}. 

\begin{table}[!ht]
\centerline{
\begin{tabular}{ccc}
\hline
\hline
HJD-2,450,000  &   RV [m~s$^{-1}$]    & Uncertainty [m~s$^{-1}$]   \\
\hline
\hline
     463.1438 &  -403.39   &   1.27 \\
     545.9798  & -390.77   &   1.29 \\
     787.1381  & -351.45   &   1.89 \\
     787.1439  & -337.14   &   1.33 \\
     838.0563  & -333.70   &   1.17 \\
     838.1109  & -329.74   &   1.10 \\
     839.1258  & -334.99   &   1.20 \\
     862.0168  & -319.81   &   1.06 \\
     862.9814  & -320.76   &   1.30 \\
     955.8422  & -313.57   &   1.45 \\
     981.7798  & -311.44   &   1.18 \\
     982.7957  & -312.89   &   1.09 \\
     1009.8359  & -302.12   &   1.13 \\
     1050.7552  & -288.92   &   1.92 \\
     1050.7572  & -290.58   &   1.29 \\
     1171.1521  & -273.81   &   1.35 \\ 
     1172.1624  & -273.77   &   1.22 \\
     1173.1611  & -277.47    &  1.31 \\ 
     1227.0404  & -267.37   &   1.28 \\
     1228.0764  & -259.05   &   1.17 \\
     1228.9999  & -262.77   &   1.37 \\
     1310.8726  & -249.58   &   1.33 \\
     1311.8551  & -243.23   &   1.27 \\
     1312.8933  & -247.37    &  1.42 \\
     1313.9960  & -249.02    &  1.40 \\
     1340.8177  & -242.25   &   1.39 \\
     1341.8676  & -248.12   &   1.49 \\
     1342.8364  & -242.09   &   1.21 \\
     1553.1269  & -205.71   &   1.55 \\
     1553.1298  & -200.00   &   1.22 \\
     1582.0826  & -196.36   &   1.34 \\
     1583.0577  & -197.94   &   1.41 \\
     1585.0085  & -196.73   &   1.20 \\
     1585.0130  & -196.74   &   1.19 \\
     1586.0218  & -205.15   &   1.34 \\
     1679.8938  & -179.28   &   1.42 \\
     1702.9103  & -175.20   &   1.28 \\
     1703.8065  & -174.94   &   1.35 \\
     1704.8793  & -177.12  &    1.49 \\
     1705.8784  & -179.77  &    1.35 \\
     1706.8757  & -175.24  &    1.43 \\
     1755.7542  & -168.56  &    1.11 \\
     \hline
\end{tabular}}
\caption{}
\end{table}

\begin{table}[!ht]
\centerline{
\begin{tabular}{ccc}
\hline
\hline
HJD-2,450,000  &   RV [m~s$^{-1}$]    & Uncertainty [m~s$^{-1}$]   \\
\hline
\hline
     1884.1599  & -141.95  &    1.41 \\
     1898.1674  & -141.82  &    1.28 \\
     1899.1728  & -137.89  &    1.37 \\
     1900.1678  & -139.61  &    1.30 \\
     1901.1771  & -138.81  &    1.27 \\
     1972.1057  & -129.95  &    1.26 \\
     1973.1304  & -127.04  &    1.33 \\
     1982.1034  & -126.24  &    1.43 \\
     2030.9064  & -125.05  &    1.43 \\
     2101.8222  & -117.06  &    1.75 \\
     2334.1177   & -67.85   &   1.84 \\
     2712.0408   &  -0.36   &   1.37 \\
     3071.9391   &  54.34   &   1.56 \\ 
    \hline
    \multicolumn{3}{c}{RV Offset} \\
    \hline
     3480.0014   & 130.10   &   0.86 \\
     3480.0026   & 131.17   &   0.62 \\
     5232.1301   & 414.79   &  0.91 \\
     5973.1039   & 529.01   &   0.95 \\
     6109.7547   &  554.22  &   0.92 \\
     6319.0020   &  584.50  &  1.10 \\
     \hline
\end{tabular}}
\caption{Doppler radial velocity measurements for HD 114174. A horizontal line indicates the division between HIRES (detector) pre-upgrade and post-upgrade data points, requiring a differential offset for analysis. Uncertainties correspond to photon-noise alone.}
\end{table}

\subsection{High-Contrast Imaging}
We acquired first-epoch high-contrast images of HD~114174 on 2011 February 22  UT using NIRC2 and the Keck II AO system \citep{wizinowich_00}. The bright ($K_s=5.20$, \citealt{skrutskie_06}) star was placed behind the 300 mas diameter coronagraphic spot. We used the angular differential imaging (ADI) technique to discriminate between residual scattered starlight and companions through point-spread-function (PSF) subtraction \citep{marois_06}. 

Raw frames were processed by flat-fielding the array, replacing bad pixels with interpolated values, and high-pass Fourier filtering. The latter step reduces the AO halo and thermal background (pedestal) while accentuating quasi-static speckles, which aids with image alignment. Frames were then multiplied by a gray-scale intensity mask (having contiguous transmission values between 0 and 1) to prevent the coronagraph from biasing image alignment. We used the Locally Optimized Combination of Images (LOCI) algorithm to increase effective contrast levels \citep{lafreniere_07}. The subtracted frames were then rotated to the standard East-North orientation using the parallactic angle to combine images and stack light from any off-axis sources. 

Upon processing the full 2011 February data set, we noticed a faint candidate companion located $\approx 0.7\arcsec$ to the south in the final reduced image. Follow-up observations were taken on 2012 February 02, 2012 May 29, and 2012 July 4 UT with the $K'$ and $J$ filters to obtain color information and determine whether the candidate is associated with the primary. Fig. 2 shows the confirmation image taken one year after the first epoch. 

We also obtained a deep ADI sequence of HD~114174 with NIRC2 in the L'-band on 2012 May 07 UT. These observations were recorded under good conditions ($\approx$0.5$\arcsec$ seeing), at low airmass (1.03-1.05), and achieved 40.5 degrees of field rotation, yet resulted in a non-detection, suggesting that HD~114174~B  may have blue colors. We derive a lower-limit to the brightness difference between HD~114174~B and its parent star of $\Delta L'>9.84$ mag. The companion was recovered at all other epochs using filters at shorter wavelengths. 

\begin{figure}[!t]
\begin{center}
\includegraphics[height=2.8in]{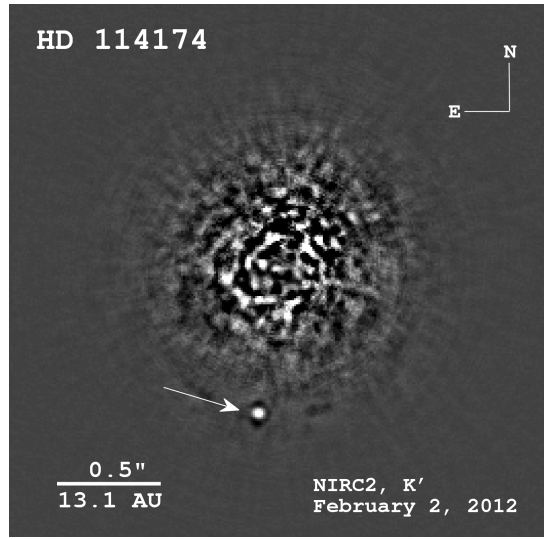} 
\caption{Confirmation image of HD~114174~B taken with NIRC2 at Keck. The $\Delta K'=10.73\pm0.12$ companion cannot be seen in individual raw frames with long exposures, but is clearly visible after PSF subtraction.} 
\end{center}\label{fig:image}
\end{figure} 

\section{PHOTOMETRY}
PSF subtraction modifies the measured flux (photometry) and location (astrometry) of faint companions that are initially buried in bright speckle noise \citep{crepp_11,pueyo_12}. We injected synthetic companions into the NIRC2 ADI data to estimate the amount of companion flux lost as a result of using the LOCI algorithm \citep{soummer_12}. Self-consistent introduction of fake companions into each frame also allows us to measure the intrinsic scatter in photometric and astrometric quantities and thus compute their uncertainty. 

\begin{table*}[!ht]
\centerline{
\begin{tabular}{lc}
\hline
\multicolumn{2}{c}{HD~114174 Properties}     \\
\hline
right ascension [J2000]            &    13 08 51.02         \\
declination [J2000]                   &    +05 12 26.06        \\
$B$                                           &    7.47                     \\
$V$                                           &     6.80                     \\
$J_{\tiny{\rm 2MASS}}$            &    $5.613\pm0.026$  \\
$K_s$                                        &   $5.202\pm0.023$  \\
$J_{\tiny{MKO}}$                      &    $5.584\pm0.027$  \\
$K_{\tiny{MKO}}$                      &   $5.189\pm0.024$  \\
d [pc]                                          &    $26.14\pm0.37$   \\
proper motion [mas/yr]              &    $84.72\pm0.59$  E \\
                                                   &   $-670.11\pm0.47$ N  \\
\hline
\multicolumn{2}{c}{Host Star}      \\
\hline
Mass [$M_{\odot}$]           &    $1.05\pm0.05$        \\
Radius [$R_{\odot}$]         &     $1.06$       \\
Luminosity [$L_{\odot}$]    &   $1.13$         \\
Age [Gyr]                            &     $4.7^{+2.3}_{-2.6}$    \\
$\mbox{[Fe/H]}$                 &      $0.07\pm0.03$      \\
log g [cm $\mbox{s}^{-2}$]    &  $4.51\pm0.06$       \\
$T_{\rm eff}$ [$K$]             &    $5781\pm44$        \\
Spectral Type                     &       G5 IV-V               \\
v sini   [km/s]                        &   $1.8\pm0.5$        \\
\hline
\multicolumn{2}{c}{Companion$^a$}  \\
\hline
$\Delta J$                 &      $10.48\pm0.11$        \\
$\Delta K$                 &      $10.75\pm0.12$       \\
$\Delta L'$                 &        $>9.84$                            \\
$J$                              &     $16.06\pm0.11$         \\
$K$                             &    $15.94\pm0.12$          \\
$L$                             &    $>14.99$                              \\
$M_J$                     &    $13.97\pm0.11$          \\
$M_K$                    &     $13.85\pm0.12$        \\
$M_{L'}$                   &      $>12.90$                            \\
$T_{\rm eff}  [K]$         &      $8200\pm4000$      \\
$\log g$ [cm/s$^2$]        &          $8.90\pm0.02$        \\
$t_c$  [Gyr]                   &        $\approx3.4$           \\
$m_{\rm dyn}$ [$M_{\odot}$]        &    $>0.260\pm0.010$               \\
$m_{\rm model}$ [$M_{\odot}$]    &    $1.15\pm0.01$  \\
\hline
\end{tabular}}
\caption{(Top) Coordinates, apparent magnitudes, distance, and proper motion of HD~114174 from the SIMBAD database. Magnitudes are from 2MASS \citep{skrutskie_06}. Filter conversions from \citet{carpenter_01} have been applied to estimate stellar MKO values. The parallax-based distance is from {\it Hipparcos} measurements using the refined data reduction of \citealt{van_leeuwen_07}. (Middle) Host star physical properties are estimated from SME using HIRES template spectra as well as theoretical isochrones \citep{valenti_fischer_05}. (Bottom) Companion magnitude difference, apparent magnitude, absolute magnitude, and companion physical properties, such as the estimated cooling age ($t_c$), mass constraint (lower-limit) from dynamics ($m_{\rm dyn}$), and mass ($m_{\rm model}$) from photometry using white dwarf evolutionary models assuming a pure hydrogen atmosphere.} 
\label{tab:starprops}
\end{table*}
%


We measured the relative brightness between HD~114174~B and its host star by comparing the flux from LOCI-processed images to unocculted images of the primary star, accounting for the difference in integration time. A fast-readout, subarray mode prevented HD~114174~A from saturating the array in unocculted frames. Subsequent reductions using fake companions and the same LOCI parameters were then performed to determine the (throughput) response of the data pipeline. The injected companions were placed at approximately the same angular separation as HD~114174~B, but with azimuthal separations sufficient to prevent artificial self-subtraction \citep{quanz_12}. Unocculted frames of the primary star informed our choice of the FWHM for fake companions. 

Measurements using two different filters, $K'_{\tiny{MKO}}$ and $J_{MKO}$, provided relative flux and color information. Optimized for the local atmospheric transmission and thermal background on Mauna Kea \citep{tokunaga_02}, the $K'_{\tiny{MKO}}$ band is our filter choice for search-mode operation. We measure flux ratios of $\Delta K'_{\tiny{MKO}}=10.73\pm0.12$ and $\Delta J_{\tiny{MKO}}=10.48\pm0.11$ (Table 3). We then convert the $J_{\tiny{\rm 2MASS}}$ and $K_s$ magnitudes of the primary star to $J_{\tiny{MKO}}$ and $K_{\tiny{MKO}}$ using the $J_{\tiny{\rm 2MASS}}-K_s$ color from SIMBAD and empirical transformations from \citet{carpenter_01}. 

To place the companion flux in the same filter basis\footnote{Central wavelength for the $K'_{\tiny{MKO}}$ and $K_{\tiny{MKO}}$ filters are $\lambda_{K'}=2.124\;\mu m$ and $\lambda_{K}=2.196\;\mu m$.}, we downloaded high-resolution spectra of various cold dwarfs from the IRTF spectral library\footnote{http://irtfweb.ifa.hawaii.edu/$\sim$spex/IRTF\_Spectral\_Library/} as well as the filter profiles from Keck Observatory\footnote{http://www2.keck.hawaii.edu/inst/nirc2old/filters.html}. IRTF library spectra from a G5 dwarf, HD 165185, served as a fiducial to compare the colors of the companion to the G5 primary, HD~114174~A. HD 165185 and HD~114174~A have $J-K_s$ colors consistent to within 0.045 mag \citep{skrutskie_06}, smaller than the 0.12 mag uncertainty in the measured companion-to-star flux difference. 

Our initial interpretation of the companion's properties led us to believe it was a brown dwarf. We thus began filter conversion calculations using IRTF spectra from two cold brown dwarfs listed in \citet{cushing_05}, SDSS J125453.90-012247.4 (T2) and 2MASS J05591915-1404489 (T4.5). We first scaled the flux of the different T-dwarf spectra by a multiplicative factor to match our $\Delta K'_{\tiny{MKO}}$ measurement, then compared the $\Delta J_{\tiny{MKO}}$ model result to our $\Delta J_{\tiny{MKO}}$ measurement. The cooler T4.5 dwarf spectrum resulted in a significantly better fit to our measurements ($\Delta J_{\tiny{MKO}}$ within $1\sigma$) than the T2 dwarf spectrum which was discrepant at the $5\sigma$ level. Using the T4.5 spectrum reference, we find $\Delta K_{\tiny{MKO}}=10.75\pm0.12$ (from the measured $\Delta K'_{\tiny{MKO}}=10.73\pm0.12$), $K_{\tiny{MKO}}=15.94\pm0.12$, and $(J-K)_{\tiny{MKO}}=0.12\pm0.16$ (Table 3). 

Both the absolute magnitude and colors of HD~114174~B are consistent with a cold brown dwarf (Fig. 3). However, as shown below (Section 5.1, Section 5.2), our dynamical analysis precludes this interpretation. Since the spectral energy distribution of cool white dwarfs and late-type brown dwarfs are both often dominated by collisionally induced $H_2$ absorption \citep{borysow_frommhold_90}, and the correction factor between the $K'_{\tiny{MKO}}$ and $K_{\tiny{MKO}}$ filters is small, particularly for dwarfs with neutral colors, we adopt the $\Delta K_{\tiny{MKO}}=10.75\pm0.12$ value for subsequent analysis. 

\begin{figure}[!t]
\begin{center}
\includegraphics[height=4.3in]{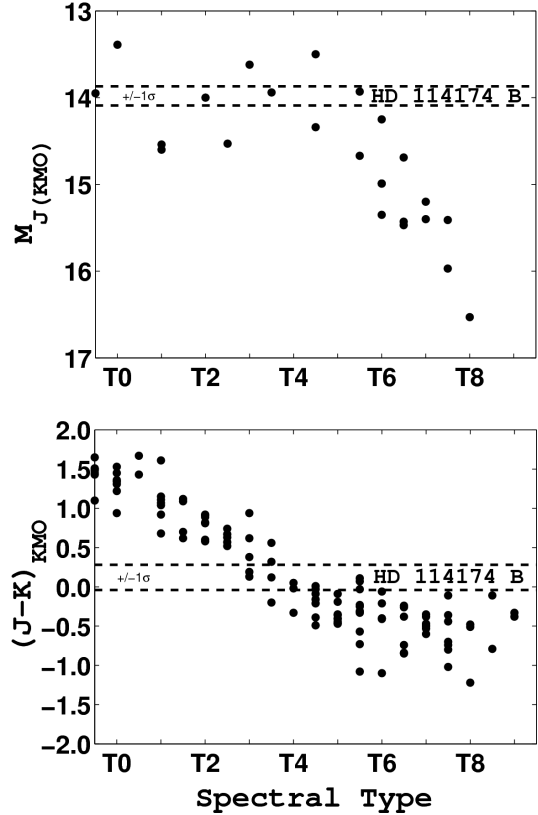} 
\caption{HD~114174~B absolute J-band magnitude and J-K color compared to cold brown dwarfs from \citet{leggett_10}. Black data points represent spectral types derived from moderate resolution spectroscopy (http://www.gemini.edu/staff/sleggett). Our $1\sigma$ measurement uncertainties from broadband photometry are shown as horizontal lines. HD~114174~B appears to be consistent with a T3.5 dwarf, but dynamical considerations indicate otherwise. Given the comparable near-infrared colors between brown dwarfs and white dwarfs, compact objects like HD~114174~B might be considered as a type of false-positive for high-contrast imaging surveys dedicated to searching for substellar companions.} 
\end{center}\label{fig:image}
\end{figure} 


\section{ASTROMETRY}
Our astrometric observations consist of four epochs taken over an 1.4 year baseline (Table~4). The proper-motion of HD~114174 is high, 675 mas $\mbox{yr}^{-1}$, allowing us to easily determine whether the companion shares the same space motion as the primary. We measured an accurate separation and position angle using the technique described in \citet{crepp_12a}. Following PSF subtraction, we fit Gaussian functions to the stellar and companion point-spread function to locate their centroids in each frame. The results from trials using numerous fake injected companions are averaged and uncertainty in the separation and position angle are initially taken as the measurement standard deviation. We then account for uncertainty in the plate scale and instrument orientation \citep{ghez_08} by propagating these errors to the final calculated position. Finally, a 5 mas uncertainty is added in quadrature to the rectilinear position of the star to account for unknown distortions introduced by the focal plane occulting mask (Q. Konopacky 2011, private communication). For companions with intensity comparable to, or fainter than, stellar quasi-static speckles using (near) Nyquist-sampled data, it is difficult to measure an astrometric position with precision much less than a single pixel \citep{marois_10_SOSIE,crepp_11}.

Fig.~4 shows multi-epoch astrometry measurements plotted against the expected motion of a distant background object. We find that HD~114174~B is clearly associated with HD~114174~A. The position of the companion has changed by only 28 mas over an 1.4 year time-frame. An unrelated background source placed at infinite distance with zero proper-motion would have moved relative to the host star by 902 mas over the same timeframe. HD~114174~B has a projected separation of $18.09\pm0.34$ AU (2012 July 4 UT) that appears to be decreasing with time. 

\begin{table*}[!t]
\centerline{
\begin{tabular}{lccccc}
\hline
\hline
  Date [UT]      &   JD-2,450,000         &     Filter    &      $\rho$ [mas]      &    Position Angle [$^{\circ}$]     &   Projected Separation [AU]   \\
\hline
\hline        
2011-02-22           &  5615.1           &             $K'$            &     $719.8\pm6.6$     &    $171.8^{\circ}\pm0.5^{\circ}$       &    $18.82\pm0.32$ \\
2012-02-02            &   5960.1            &            $K'$          &      $701.1\pm5.0$    &    $172.1^{\circ}\pm0.4^{\circ}$       &    $18.33\pm0.29$ \\
2012-05-29            &   6076.8            &            $K'$          &     $695.8\pm5.8$      &    $170.5^{\circ}\pm0.6^{\circ}$       &    $18.19\pm0.25$  \\
2012-07-04            &   6112.8            &            $ J$          &     $692.1\pm8.7$      &    $172.2^{\circ}\pm0.8^{\circ}$       &    $18.09\pm0.34$ \\
\hline
\hline
\end{tabular}}
\caption{Summary of astrometric measurements.}
\label{tab:astrometry}
\end{table*}

\begin{figure}[!t]
\begin{center}
\includegraphics[height=2.8in]{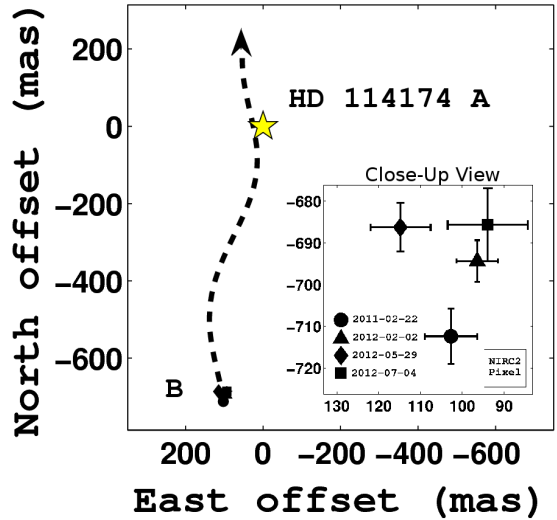} 
\caption{Astrometric measurements demonstrating that HD 114174 B is co-moving with its host star. Location of the primary defines the coordinate system and is denoted by a star. The dashed curve shows the path that a distant background object would execute from 2011 February 22 UT through 2012 July 4 UT, accounting for proper-motion and parallactic motion.} 
\end{center}\label{fig:astrometry}
\end{figure} 

\section{COMPANION PROPERTIES}
\subsection{Mass Lower-limit from Dynamics}
A lower-limit to the companion mass may be determined from the RV acceleration and direct imaging angular separation \citep{torres_99,liu_02}. Although the HD~114174 Doppler signal already shows a hint of curvature (Fig. 1), we do not compute the full range of possible masses because the astrometry has yet to show significant orbital motion, which is {\it required} to place interesting constraints on orbital elements and the companion mass \citep{kalas_08,currie_11_betapic,crepp_12a,crepp_12b}. Instead, we evaluate the local slope of the RV time series (see also Section 5.4). We have obtained a recent Doppler RV observation (2013 January 26) to facilitate this calculation. Using an MCMC simulation to fit the instantaneous RV slope, we find a best-fitting Doppler acceleration of $57.4\pm0.4$ m$\:\rm{s^{-1} yr^{-1}}$, a value similar to that found in Section 2.1 when considering the full RV data set because curvature in the RV time-series is still subtle. This measurement translates to a minimum dynamical mass of $0.260\pm0.010M_{\odot}$, thus precluding the interpretation that HD~114174~B is substellar, despite its low luminosity and neutral colors. 

\subsection{White Dwarf Interpretation}   
White dwarfs are a common outcome of the evolutionary process for low-mass stars and are thus expected to comprise a non-negligible fraction of stellar companions \citep{holberg_98,breton_12}. With a sample of more than 100 targets, it is therefore not surprising that the TRENDS program would discover a compact object orbiting a nearby star. Indeed, it seems the only self-consistent interpretation of our photometric and dynamical measurements is that HD~114174~B is a white dwarf companion. Such serendipitous discoveries might be considered as direct imaging false-positives when searching for exoplanets and brown dwarfs \citep{zurlo_13}.

Despite the prevalence of white dwarfs throughout the galaxy \citep{oppenheimer_01}, relatively few benchmark compact objects have been discovered \citep{muirhead_13} and only several nearby companions with precise parallax measurements, such as Sirius B and Procyon B, have been imaged directly \citep{liebert_05}. Similar to field brown dwarfs, most detailed white dwarf studies involve age calibration based on stellar cluster membership \citep{williams_04}, and those with reliable ages rarely have simultaneous dynamical mass estimates \citep{girard_00,bond_09}. HD~114174 thus represents a convenient system for studying the intricacies of post-main-sequence stellar evolution at low mass. 

Without spectra, we cannot determine the chemical composition of HD~114174~B. Instead, we proceed by considering two different canonical cases: pure hydrogen and pure helium white dwarf atmospheres. By fitting theoretical models to our $J$ and $K$ observations, we can constrain the effective temperature, mass, and cooling age of HD~114174~B under reasonable assumptions about its composition. 

\subsection{Synthetic Photometry \& Fitting}
For pure hydrogen atmospheres and synthetic spectra, we use the theoretical models of \citet{tremblay_11}.\footnote{http://www.astro.umontreal.ca/$\sim$bergeron/CoolingModels/} The models include improved calculations for Stark broadening of hydrogen lines \citep{tremblay_09} and non-ideal perturbations, described within the occupation probability formalism of \citet{hummer_88}, directly inside the line profile calculations. The models also include Lyman alpha profile calculations from \citet{kowalski_06}. Convective energy transport is taken into account following the revised ML2/$\alpha$ = 0.8 prescription of mixing-length theory from \citet{tremblay_10}. Details regarding our helium atmosphere models are provided in \citet{bergeron_11}. Model grids cover a range in effective temperature between $T_{\rm eff}~= 1500$ K and 40,000 K in steps varying from 500~K to 5000~K. The surface gravity ranges from $7.0 \leq \log g \leq 9.0$ by steps of 0.5 dex. Synthetic colors are obtained using the procedure of \citet{holberg_06} based on Vega fluxes taken from \citet{bohlin_04}.

\begin{figure}[!t]
\begin{center}
\includegraphics[scale=0.52,bb= 150 210 420 560]{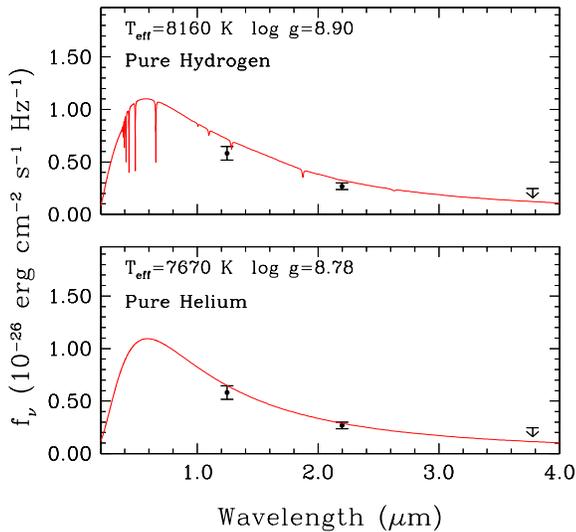}
\caption{White dwarf model fits to our JKL' photometry assuming pure hydrogen (top) and pure helium (bottom) atmospheres. The observed JK fluxes are represented by errorbars. Model monochromatic fluxes are shown as the solid red line. Our measurements yield physical parameter estimates as listed in Table 3. The L'-band non-detection from Keck is shown as an upper-limit.} 
\end{center}\label{fig:wdmodel}
\end{figure} 

The method used to fit photometric data is similar to that described in \citet{bergeron_01}, which we briefly summarize. We first transform the magnitudes in each bandpass into observed average fluxes $f_{\lambda,m}$ using the following equation:
\begin{equation}
  m = -2.5 \log f_{\lambda,m} + c_{m},
\label{eq:1}
\end{equation}
\noindent where
\begin{equation}
  f_{\lambda,m} = \frac{\int^{\infty}_{0}f_{\lambda}S_{m}(\lambda)d\lambda}{\int^{\infty}_{0}S_{m}(\lambda)d\lambda}.
\label{eq:2}
\end{equation}

\noindent The transmission functions $S_{m}(\lambda)$ along with the constants $c_{m}$ for each bandpass are described at length in \citet{holberg_06} and references therein. To make use of all photometric measurements simultaneously, we convert measured magnitudes into observed fluxes using equation (\ref{eq:1}) and compare the resulting energy distributions with those predicted from our model atmosphere calculations. Model fluxes are also averaged over filter bandpasses by substituting $f_{\lambda}$ in equation (\ref{eq:2}) for the monochromatic Eddington flux $H_{\lambda}$. The average observed fluxes $f_{\lambda,m}$ and model fluxes $H_{\lambda,m}$, which depend on $T_{\rm eff}$, $\log g$ and $N$(He)/$N$(H), are related by
\begin{equation}
  f_{\lambda,m} = 4\pi(R/d)^{2}H_{\lambda,m} 
\label{eq:3}
\end{equation}
\noindent where $R/d$ is the ratio of the radius of the star to its distance from Earth. 

Our fitting procedure relies on the nonlinear, least-squares, steepest descent method of Levenberg-Marquardt. The value of $\chi^{2}$ is taken as the sum over all bandpasses of the difference between both sides of equation (\ref{eq:3}), properly weighted by the corresponding observational uncertainties. In our fitting procedure, we consider only $T_{\rm eff}$~and the solid angle, $R/d$, as free parameters. We begin by guessing a value for the surface gravity ($\log g\;=\;8$) and determine the corresponding effective temperature and solid angle. Combined with the distance,  $d$, obtained from the trigonometric parallax measurement, this yields directly the radius of the star $R$. The radius is then converted into mass using evolutionary models similar to those described in \citet{fontaine_01} but with C/O cores, $q$(He) $\equiv$ $\log M_{He}/M_{\star} = 10^{-2}$ and $q(H) = 10^{-4}$, which are representative of hydrogen atmosphere white dwarfs, and $q(He) = 10^{-2}$ and $q(H) = 10^{-10}$, which are representative of helium atmosphere white dwarfs. In general, the $\log g$ value obtained from the inferred mass and radius ($g = GM/R^{2}$) will be different from our initial guess of $\log g=8$, and the fitting procedure is thus repeated until an internal consistency in $\log g$ is reached.

Assuming a hydrogen-dominated atmosphere, we find: $T_{\rm eff} =  8200\pm4000$ K and $\log g=8.90\pm0.02$, which corresponds to a cooling age of $t_c\approx3.4$ Gyr and model-dependent mass $m_{\rm model}=1.15\pm0.01M_{\odot}$. Assuming a Helium-dominated atmosphere, we find: $T_{\rm eff}=7700\pm4200$ K, $\log g=8.78\pm0.02$, $t_c\approx3.5$ Gyr, and $m_{\rm model}=1.08\pm0.01M_{\odot}$. With negligible differences between the physical parameter determinations when considering both models, we adopt the pure-H model values (Table 3). An $m=1.1M_{\odot}$ remnant corresponds to an $\approx$$6.2M_{\odot}-6.9M_{\odot}$ progenitor mass \citep{kalirai_08,williams_09}. Thus, HD~114174~B spent only a short period of time ($\approx 50$ Myr) on the main-sequence. 

\subsection{Period Range from Dynamics and Estimated Companion Mass}   
With an estimate of the mass of HD~114174~B, we can constrain the system orbital period using dynamics. We first determine the true (physical) separation of HD~114174~B from HD~114174~A using the instantaneous Doppler acceleration ($57.4\pm0.4$ m$\:\rm{s^{-1} yr^{-1}}$) and measured angular separation ($692.1\pm8.7$ mas) from geometry (e.g., \citealt{howard_10}). We find that HD~114174~B is presently $59.8\pm0.4$ AU from its host star, much further than the projected separation ($18.09\pm0.34$ AU). The true anomaly cannot be determined from the current data set. However, assuming the orbit configuration is near apastron or periastron, we can constrain the semimajor axis and thus orbital period. We find that the period is between 154--881 years for eccentricities in the range $0 \leq e \leq0.5$. The period lower-limit corresponds to an expected RV semi-amplitude of $K\approx4400$ m/s assuming an edge-on orbit, consistent with Fig. 1.

Our numerical simulations indicate that $\approx5-10$\% of an orbit cycle is nominally required following a direct imaging discovery to place interesting constraints on all six orbit elements and the companion mass using dynamics for astrometric and Doppler RV signal-to-noise levels comparable to those of HD~114174~B \citep{crepp_2012AAS}. Thus, we may be able to determine the system orbit and mass of HD~114174~B as early as $\approx$8 years from now with sufficient observing cadence. 

\section{SUMMARY \& DISCUSSION}
We report the discovery of a faint companion orbiting the nearby G5 star HD~114174. Initially identified as a promising high-contrast imaging target due to a measured long-term RV acceleration, we have used NIRC2 at Keck to obtain direct images of the source that is responsible for causing the Doppler trend. Multi-epoch detections in two different filters demonstrate association through common proper-motion over an 1.4 year time frame. The RV data spans 16.0 years and shows curvature (a change in the acceleration). Since HD~114174 has a precise parallax measurement, it will be possible to place strong constraints on the companion orbit (all six parameters) and mass as soon as the direct imaging astrometry shows curvature. 

Our broadband photometric measurements indicate that the companion has neutral, or possibly blue colors ($J-K=0.12\pm0.16$) suggesting interesting atmospheric physics. Deep ADI observations in the {\it L'}-band under good seeing conditions at Keck resulted in a non-detection, further supporting the interpretation of a decreasing NIR spectral-energy distribution with increasing wavelength. HD~114174~B has an absolute magnitude ($M_J=13.97\pm0.11$) and colors consistent with a T3-dwarf. However, dynamical considerations preclude the interpretation of a substellar companion. We calculate a firm minimum mass of $m>0.260\pm0.010M_{\odot}$ using presently available RV and imaging data. 

HD~114174~B appears to be a white dwarf. Using theoretical models to fit our photometry and assuming a pure hydrogen atmosphere, we estimate $T_{\rm eff} =  8200\pm4000$ K and $\log g=8.90\pm0.02$, which corresponds to a cooling age of $t_c\approx3.4$ Gyr and remnant mass $m_{\rm model}=1.15\pm0.01M_{\odot}$, consistent with the $4.7^{+2.3}_{-2.6}$ Gyr age estimate of the system based on isochronal analysis of HD~114174~A, and also consistent with the dynamical mass lower limit. Calculations assuming a Helium-rich atmosphere yield similar results. 

Uncertainty in $T_{\rm eff}$ is presently limited by the precision of our $J$ and $K$ photometry and current wavelength coverage. Further, the NIR colors of cool white dwarfs are known to exhibit an inversion (from having red colors to blue colors as $T_{\rm eff}$ decreases) as the result of collisionally-induced H$_2$ absorption. Thus, there exists a degeneracy in $T_{\rm eff}$ for $J-K\approx0.0$, possibly complicating the interpretation of HD~114174~B's spectrophotometric signal at the lowest temperatures currently allowed by our models. For example, the white dwarf companion Gliese 86 B ($J-K=1.0\pm0.3$) was initially characterized as a brown dwarf \citep{els_01,mugrauer_05}. Deeper L'-band observations obtained with the LBT \citep{skrutskie_10,skemer_12} and direct NIR spectroscopy measurements using an integral-field unit \citep{hinkley_11_PASP} will allow us to refine HD~114174~B's physical properties. Furthermore, it may be possible to study HD~114174~B at shorter wavelengths using HST in a fashion similar to that of \citet{barstow_05} who obtained a spectrum of Sirius B with STIS. 

HD~114174 A is a marginally evolved Sun-like star located at $d=26.14\pm0.37$ pc. Thus, it may be possible to measure its radius directly using CHARA interferometry, placing tighter constraints on the primary star's mass and hence the system age and also secondary mass \citep{boyajian_09}. A similar analysis has been performed for the benchmark brown dwarf HR~7672 B \citep{crepp_12a}. In the case of a white dwarf, a more accurate mass and age will likewise constrain physical properties of the progenitor. The HD~114174 system thus represents a useful testbed for studying white dwarf initial-to-final mass ratios at a level of detail comparable to Sirius and Procyon. Given the anticipated sensitivity of forth-coming high-contrast imaging programs that use ``extreme" AO, it is likely that additional benchmark white dwarf companions will be (accidentally) uncovered in the near future. A more complete census of Sirius-like systems may ultimately be undertaken with the GAIA space mission using precision astrometry.

\section{ACKNOWLEDGEMENTS}
We thank Sasha Hinkley for trading NIRC2 observing time that allowed us to acquire observations of HD~114174~B in a complementary bandpass, demonstrating its unusual colors, and Mike Liu for pointing out reference to the Gliese 86 system which shares a similar story as HD~114174~B for being a white dwarf initially characterized as a brown dwarf. The referee, John Subasavage, provided helpful comments that improved the clarity of our manuscript. This research has made use of the SIMBAD database, operated at CDS, Strasbourg, France. The TRENDS high-contrast imaging program is supported by NASA Origins grant NNX13AB03G. JAJ is supported by generous grants from the David and Lucile Packard Foundation and the Alfred P. Sloan Foundation. 

Data presented herein were obtained at the W.M. Keck Observatory, which is operated as a scientific partnership among the California Institute of Technology, the University of California and the National Aeronautics and Space Administration. The Observatory was made possible by the generous financial support of the W.M. Keck Foundation. The Center for Exoplanets and Habitable Worlds is supported by the Pennsylvania State University, the Eberly College of Science, and the Pennsylvania Space Grant Consortium.

\begin{small}
\bibliographystyle{jtb}
\bibliography{ms.bib}

\begin{thebibliography}{}

\bibitem[\protect\astroncite{{Barstow} et~al.}{2005}]{barstow_05}
{Barstow}, M.~A., {Bond}, H.~E., {Holberg}, J.~B., {Burleigh}, M.~R., {Hubeny},
  I., and {Koester}, D. (2005) ,
\newblock {\em \mnras} {\bf 362}, 1134

\bibitem[\protect\astroncite{{Bergeron} et~al.}{2001}]{bergeron_01}
{Bergeron}, P., {Leggett}, S.~K., and {Ruiz}, M.~T. (2001) ,
\newblock {\em \apjs} {\bf 133}, 413

\bibitem[\protect\astroncite{{Bergeron} et~al.}{2011}]{bergeron_11}
{Bergeron}, P., {Wesemael}, F., {Dufour}, P., {Beauchamp}, A., {Hunter}, C.,
  {Saffer}, R.~A., {Gianninas}, A., {Ruiz}, M.~T., {Limoges}, M.-M., {Dufour},
  P., {Fontaine}, G., and {Liebert}, J. (2011) ,
\newblock {\em \apj} {\bf 737}, 28

\bibitem[\protect\astroncite{{Bohlin} and {Gilliland}}{2004}]{bohlin_04}
{Bohlin}, R.~C. and {Gilliland}, R.~L. (2004) ,
\newblock {\em \aj} {\bf 127}, 3508

\bibitem[\protect\astroncite{{Bond}}{2009}]{bond_09}
{Bond}, H.~E. (2009) ,
\newblock {\em Journal of Physics Conference Series} {\bf 172(1)}, 012029

\bibitem[\protect\astroncite{{Borysow} and
  {Frommhold}}{1990}]{borysow_frommhold_90}
{Borysow}, A. and {Frommhold}, L. (1990) ,
\newblock {\em \apjl} {\bf 348}, L41

\bibitem[\protect\astroncite{{Boyajian} et~al.}{2009}]{boyajian_09}
{Boyajian}, T.~S., {McAlister}, H.~A., {Cantrell}, J.~R., {Gies}, D.~R.,
  {Brummelaar}, T.~A.~t., {Farrington}, C., {Goldfinger}, P.~J., {Sturmann},
  L., {Sturmann}, J., {Turner}, N.~H., and {Ridgway}, S. (2009) ,
\newblock {\em \apj} {\bf 691}, 1243

\bibitem[\protect\astroncite{{Breton} et~al.}{2012}]{breton_12}
{Breton}, R.~P., {Rappaport}, S.~A., {van Kerkwijk}, M.~H., and {Carter}, J.~A.
  (2012) ,
\newblock {\em \apj} {\bf 748}, 115

\bibitem[\protect\astroncite{{Carpenter}}{2001}]{carpenter_01}
{Carpenter}, J.~M. (2001) ,
\newblock {\em \aj} {\bf 121}, 2851

\bibitem[\protect\astroncite{{Crepp} et~al.}{2010}]{crepp_10}
{Crepp}, J., {Serabyn}, E., {Carson}, J., {Ge}, J., and {Kravchenko}, I. (2010)
  ,
\newblock {\em \apj} {\bf 715}, 1533

\bibitem[\protect\astroncite{{Crepp} and {Johnson}}{2011}]{crepp_johnson_11}
{Crepp}, J.~R. and {Johnson}, J.~A. (2011) ,
\newblock In {\em American Astronomical Society Meeting Abstracts}, Vol. 217 of
  {\em American Astronomical Society Meeting Abstracts}, p. 302.05

\bibitem[\protect\astroncite{{Crepp} et~al.}{2012a}]{crepp_12a}
{Crepp}, J.~R., {Johnson}, J.~A., {Fischer}, D.~A., {Howard}, A.~W., {Marcy},
  G.~W., {Wright}, J.~T., {Isaacson}, H., {Boyajian}, T., {von Braun}, K.,
  {Hillenbrand}, L.~A., {Hinkley}, S., {Carpenter}, J.~M., and {Brewer}, J.~M.
  (2012a) ,
\newblock {\em \apj} {\bf 751}, 97

\bibitem[\protect\astroncite{{Crepp} et~al.}{2012b}]{crepp_12b}
{Crepp}, J.~R., {Johnson}, J.~A., {Howard}, A.~W., {Marcy}, G.~W., {Fischer},
  D.~A., {Hillenbrand}, L.~A., {Yantek}, S.~M., {Delaney}, C.~R., {Wright},
  J.~T., {Isaacson}, H.~T., and {Montet}, B.~T. (2012b) ,
\newblock {\em \apj} {\bf 761}, 39

\bibitem[\protect\astroncite{{Crepp} et~al.}{2012c}]{crepp_13a}
{Crepp}, J.~R., {Johnson}, J.~A., {Howard}, A.~W., {Marcy}, G.~W., {Fischer},
  D.~A., {Yantek}, S.~M., {Wright}, J.~T., {Isaacson}, H., and {Feng}, Y.
  (2012c) ,
\newblock {\em ArXiv e-prints}

\bibitem[\protect\astroncite{{Crepp} et~al.}{2012d}]{crepp_2012AAS}
{Crepp}, J.~R., {Johnson}, J.~A., and {Planet Search}, C. (2012d) ,
\newblock In {\em American Astronomical Society Meeting Abstracts 220}, Vol.
  220 of {\em American Astronomical Society Meeting Abstracts}, p. 120.09

\bibitem[\protect\astroncite{{Crepp} et~al.}{2011}]{crepp_11}
{Crepp}, J.~R., {Pueyo}, L., {Brenner}, D., {Oppenheimer}, B.~R., {Zimmerman},
  N., {Hinkley}, S., {Parry}, I., {King}, D., {Vasisht}, G., {Beichman}, C.,
  {Hillenbrand}, L., {Dekany}, R., {Shao}, M., {Burruss}, R., {Roberts}, L.~C.,
  {Bouchez}, A., {Roberts}, J., and {Soummer}, R. (2011) ,
\newblock {\em \apj} {\bf 729}, 132

\bibitem[\protect\astroncite{{Currie} et~al.}{2011}]{currie_11_betapic}
{Currie}, T., {Thalmann}, C., {Matsumura}, S., {Madhusudhan}, N., {Burrows},
  A., and {Kuchner}, M. (2011) ,
\newblock {\em \apjl} {\bf 736}, L33+

\bibitem[\protect\astroncite{{Cushing} et~al.}{2005}]{cushing_05}
{Cushing}, M.~C., {Rayner}, J.~T., and {Vacca}, W.~D. (2005) ,
\newblock {\em \apj} {\bf 623}, 1115

\bibitem[\protect\astroncite{{Els} et~al.}{2001}]{els_01}
{Els}, S.~G., {Sterzik}, M.~F., {Marchis}, F., {Pantin}, E., {Endl}, M., and
  {K{\"u}rster}, M. (2001) ,
\newblock {\em \aap} {\bf 370}, L1

\bibitem[\protect\astroncite{{Fontaine} et~al.}{2001}]{fontaine_01}
{Fontaine}, G., {Brassard}, P., and {Bergeron}, P. (2001) ,
\newblock {\em \pasp} {\bf 113}, 409

\bibitem[\protect\astroncite{{Ghez} et~al.}{2008}]{ghez_08}
{Ghez}, A.~M., {Salim}, S., {Weinberg}, N.~N., {Lu}, J.~R., {Do}, T., {Dunn},
  J.~K., {Matthews}, K., {Morris}, M.~R., {Yelda}, S., {Becklin}, E.~E.,
  {Kremenek}, T., {Milosavljevic}, M., and {Naiman}, J. (2008) ,
\newblock {\em \apj} {\bf 689}, 1044

\bibitem[\protect\astroncite{{Girard} et~al.}{2000}]{girard_00}
{Girard}, T.~M., {Wu}, H., {Lee}, J.~T., {Dyson}, S.~E., {van Altena}, W.~F.,
  {Horch}, E.~P., {Gilliland}, R.~L., {Schaefer}, K.~G., {Bond}, H.~E.,
  {Ftaclas}, C., {Brown}, R.~H., {Toomey}, D.~W., {Shipman}, H.~L.,
  {Provencal}, J.~L., and {Pourbaix}, D. (2000) ,
\newblock {\em \aj} {\bf 119}, 2428

\bibitem[\protect\astroncite{{Hinkley} et~al.}{2011}]{hinkley_11_PASP}
{Hinkley}, S., {Oppenheimer}, B.~R., {Zimmerman}, N., {Brenner}, D., {Parry},
  I.~R., {Crepp}, J.~R., {Vasisht}, G., {Ligon}, E., {King}, D., {Soummer}, R.,
  {Sivaramakrishnan}, A., {Beichman}, C., {Shao}, M., {Roberts}, L.~C.,
  {Bouchez}, A., {Dekany}, R., {Pueyo}, L., {Roberts}, J.~E., {Lockhart}, T.,
  {Zhai}, C., {Shelton}, C., and {Burruss}, R. (2011) ,
\newblock {\em \pasp} {\bf 123}, 74

\bibitem[\protect\astroncite{{Holberg}}{2009}]{holberg_09}
{Holberg}, J.~B. (2009) ,
\newblock {\em Journal of Physics Conference Series} {\bf 172(1)}, 012022

\bibitem[\protect\astroncite{{Holberg} et~al.}{1998}]{holberg_98}
{Holberg}, J.~B., {Barstow}, M.~A., {Bruhweiler}, F.~C., {Cruise}, A.~M., and
  {Penny}, A.~J. (1998) ,
\newblock {\em \apj} {\bf 497}, 935

\bibitem[\protect\astroncite{{Holberg} and {Bergeron}}{2006}]{holberg_06}
{Holberg}, J.~B. and {Bergeron}, P. (2006) ,
\newblock {\em \aj} {\bf 132}, 1221

\bibitem[\protect\astroncite{{Howard} et~al.}{2010}]{howard_10}
{Howard}, A.~W., {Johnson}, J.~A., {Marcy}, G.~W., {Fischer}, D.~A., {Wright},
  J.~T., {Bernat}, D., {Henry}, G.~W., {Peek}, K.~M.~G., {Isaacson}, H.,
  {Apps}, K., {Endl}, M., {Cochran}, W.~D., {Valenti}, J.~A., {Anderson}, J.,
  and {Piskunov}, N.~E. (2010) ,
\newblock {\em \apj} {\bf 721}, 1467

\bibitem[\protect\astroncite{{Hummer} and {Mihalas}}{1988}]{hummer_88}
{Hummer}, D.~G. and {Mihalas}, D. (1988) ,
\newblock {\em \apj} {\bf 331}, 794

\bibitem[\protect\astroncite{{Kalas} et~al.}{2008}]{kalas_08}
{Kalas}, P., {Graham}, J.~R., {Chiang}, E., {Fitzgerald}, M.~P., {Clampin}, M.,
  {Kite}, E.~S., {Stapelfeldt}, K., {Marois}, C., and {Krist}, J. (2008) ,
\newblock {\em Science} {\bf 322}, 1345

\bibitem[\protect\astroncite{{Kalirai} et~al.}{2008}]{kalirai_08}
{Kalirai}, J.~S., {Hansen}, B.~M.~S., {Kelson}, D.~D., {Reitzel}, D.~B.,
  {Rich}, R.~M., and {Richer}, H.~B. (2008) ,
\newblock {\em \apj} {\bf 676}, 594

\bibitem[\protect\astroncite{{Kowalski} and {Saumon}}{2006}]{kowalski_06}
{Kowalski}, P.~M. and {Saumon}, D. (2006) ,
\newblock {\em \apjl} {\bf 651}, L137

\bibitem[\protect\astroncite{{Lafreni{\`e}re} et~al.}{2007}]{lafreniere_07}
{Lafreni{\`e}re}, D., {Doyon}, R., {Marois}, C., {Nadeau}, D., {Oppenheimer},
  B.~R., {Roche}, P.~F., {Rigaut}, F., {Graham}, J.~R., {Jayawardhana}, R.,
  {Johnstone}, D., {Kalas}, P.~G., {Macintosh}, B., and {Racine}, R. (2007) ,
\newblock {\em \apj} {\bf 670}, 1367

\bibitem[\protect\astroncite{{Leggett} et~al.}{2010}]{leggett_10}
{Leggett}, S.~K., {Burningham}, B., {Saumon}, D., {Marley}, M.~S., {Warren},
  S.~J., {Smart}, R.~L., {Jones}, H.~R.~A., {Lucas}, P.~W., {Pinfield}, D.~J.,
  and {Tamura}, M. (2010) ,
\newblock {\em \apj} {\bf 710}, 1627

\bibitem[\protect\astroncite{{Liebert} et~al.}{2005}]{liebert_05}
{Liebert}, J., {Young}, P.~A., {Arnett}, D., {Holberg}, J.~B., and {Williams},
  K.~A. (2005) ,
\newblock {\em \apjl} {\bf 630}, L69

\bibitem[\protect\astroncite{{Liu} et~al.}{2002}]{liu_02}
{Liu}, M.~C., {Fischer}, D.~A., {Graham}, J.~R., {Lloyd}, J.~P., {Marcy},
  G.~W., and {Butler}, R.~P. (2002) ,
\newblock {\em \apj} {\bf 571}, 519

\bibitem[\protect\astroncite{{Marcy} and {Butler}}{1992}]{marcy_butler_92}
{Marcy}, G.~W. and {Butler}, R.~P. (1992) ,
\newblock {\em \pasp} {\bf 104}, 270

\bibitem[\protect\astroncite{{Marois} et~al.}{2006}]{marois_06}
{Marois}, C., {Lafreni{\`e}re}, D., {Doyon}, R., {Macintosh}, B., and {Nadeau},
  D. (2006) ,
\newblock {\em \apj} {\bf 641}, 556

\bibitem[\protect\astroncite{{Marois} et~al.}{2010}]{marois_10_SOSIE}
{Marois}, C., {Macintosh}, B., and {V{\'e}ran}, J.-P. (2010) ,
\newblock In {\em Society of Photo-Optical Instrumentation Engineers (SPIE)
  Conference Series}, Vol. 7736 of {\em Society of Photo-Optical
  Instrumentation Engineers (SPIE) Conference Series}

\bibitem[\protect\astroncite{{Mugrauer} and
  {Neuh{\"a}user}}{2005}]{mugrauer_05}
{Mugrauer}, M. and {Neuh{\"a}user}, R. (2005) ,
\newblock {\em \mnras} {\bf 361}, L15

\bibitem[\protect\astroncite{{Muirhead} et~al.}{2013}]{muirhead_13}
{Muirhead}, P.~S., {Vanderburg}, A., {Shporer}, A., {Becker}, J., {Swift},
  J.~J., {Lloyd}, J.~P., {Fuller}, J., {Zhao}, M., {Hinkley}, S., {Pineda},
  J.~S., {Bottom}, M., {Howard}, A.~W., {von Braun}, K., {Boyajian}, T.~S.,
  {Law}, N., {Baranec}, C., {Riddle}, R., {Ramaprakash}, A.~N., {Tendulkar},
  S.~P., {Bui}, K., {Burse}, M., {Chordia}, P., {Das}, H., {Dekany}, R.,
  {Punnadi}, S., and {Johnson}, J.~A. (2013) ,
\newblock {\em \apj} {\bf 767}, 111

\bibitem[\protect\astroncite{{Oppenheimer} et~al.}{2001}]{oppenheimer_01}
{Oppenheimer}, B.~R., {Hambly}, N.~C., {Digby}, A.~P., {Hodgkin}, S.~T., and
  {Saumon}, D. (2001) ,
\newblock {\em Science} {\bf 292}, 698

\bibitem[\protect\astroncite{{Pueyo} et~al.}{2012}]{pueyo_12}
{Pueyo}, L., {Crepp}, J.~R., {Vasisht}, G., {Brenner}, D., {Oppenheimer},
  B.~R., {Zimmerman}, N., {Hinkley}, S., {Parry}, I., {Beichman}, C.,
  {Hillenbrand}, L., {Roberts}, L.~C., {Dekany}, R., {Shao}, M., {Burruss}, R.,
  {Bouchez}, A., {Roberts}, J., and {Soummer}, R. (2012) ,
\newblock {\em \apjs} {\bf 199}, 6

\bibitem[\protect\astroncite{{Quanz} et~al.}{2012}]{quanz_12}
{Quanz}, S.~P., {Crepp}, J.~R., {Janson}, M., {Avenhaus}, H., {Meyer}, M.~R.,
  and {Hillenbrand}, L.~A. (2012) ,
\newblock {\em \apj} {\bf 754}, 127

\bibitem[\protect\astroncite{{Skemer} et~al.}{2012}]{skemer_12}
{Skemer}, A.~J., {Hinz}, P.~M., {Esposito}, S., {Burrows}, A., {Leisenring},
  J., {Skrutskie}, M., {Desidera}, S., {Mesa}, D., {Arcidiacono}, C.,
  {Mannucci}, F., {Rodigas}, T.~J., {Close}, L., {McCarthy}, D., {Kulesa}, C.,
  {Agapito}, G., {Apai}, D., {Argomedo}, J., {Bailey}, V., {Boutsia}, K.,
  {Briguglio}, R., {Brusa}, G., {Busoni}, L., {Claudi}, R., {Eisner}, J.,
  {Fini}, L., {Follette}, K.~B., {Garnavich}, P., {Gratton}, R., {Guerra},
  J.~C., {Hill}, J.~M., {Hoffmann}, W.~F., {Jones}, T., {Krejny}, M., {Males},
  J., {Masciadri}, E., {Meyer}, M.~R., {Miller}, D.~L., {Morzinski}, K.,
  {Nelson}, M., {Pinna}, E., {Puglisi}, A., {Quanz}, S.~P., {Quiros-Pacheco},
  F., {Riccardi}, A., {Stefanini}, P., {Vaitheeswaran}, V., {Wilson}, J.~C.,
  and {Xompero}, M. (2012) ,
\newblock {\em \apj} {\bf 753}, 14

\bibitem[\protect\astroncite{{Skrutskie} et~al.}{2006}]{skrutskie_06}
{Skrutskie}, M.~F., {Cutri}, R.~M., {Stiening}, R., {Weinberg}, M.~D.,
  {Schneider}, S., {Carpenter}, J.~M., {Beichman}, C., {Capps}, R., {Chester},
  T., {Elias}, J., {Huchra}, J., {Liebert}, J., {Lonsdale}, C., {Monet}, D.~G.,
  {Price}, S., {Seitzer}, P., {Jarrett}, T., {Kirkpatrick}, J.~D., {Gizis},
  J.~E., {Howard}, E., {Evans}, T., {Fowler}, J., {Fullmer}, L., {Hurt}, R.,
  {Light}, R., {Kopan}, E.~L., {Marsh}, K.~A., {McCallon}, H.~L., {Tam}, R.,
  {Van Dyk}, S., and {Wheelock}, S. (2006) ,
\newblock {\em \aj} {\bf 131}, 1163

\bibitem[\protect\astroncite{{Skrutskie} et~al.}{2010}]{skrutskie_10}
{Skrutskie}, M.~F., {Jones}, T., {Hinz}, P., {Garnavich}, P., {Wilson}, J.,
  {Nelson}, M., {Solheid}, E., {Durney}, O., {Hoffmann}, W., {Vaitheeswaran},
  V., {McMahon}, T., {Leisenring}, J., and {Wong}, A. (2010) ,
\newblock In {\em Society of Photo-Optical Instrumentation Engineers (SPIE)
  Conference Series}, Vol. 7735 of {\em Society of Photo-Optical
  Instrumentation Engineers (SPIE) Conference Series}

\bibitem[\protect\astroncite{{Soummer} et~al.}{2012}]{soummer_12}
{Soummer}, R., {Pueyo}, L., and {Larkin}, J. (2012) ,
\newblock {\em \apjl} {\bf 755}, L28

\bibitem[\protect\astroncite{{Tokovinin}}{2004}]{tokovinin_04}
{Tokovinin}, A. (2004) ,
\newblock In {\em Revista Mexicana de Astronomia y Astrofisica Conference
  Series}.  (C. {Allen} and C. {Scarfe} eds.), Vol.~21 of {\em Revista Mexicana
  de Astronomia y Astrofisica Conference Series}, pp. 7--14

\bibitem[\protect\astroncite{{Tokunaga} et~al.}{2002}]{tokunaga_02}
{Tokunaga}, A.~T., {Simons}, D.~A., and {Vacca}, W.~D. (2002) ,
\newblock {\em \pasp} {\bf 114}, 180

\bibitem[\protect\astroncite{{Torres}}{1999}]{torres_99}
{Torres}, G. (1999) ,
\newblock {\em \pasp} {\bf 111}, 169

\bibitem[\protect\astroncite{{Tremblay} and {Bergeron}}{2009}]{tremblay_09}
{Tremblay}, P.-E. and {Bergeron}, P. (2009) ,
\newblock {\em \apj} {\bf 696}, 1755

\bibitem[\protect\astroncite{{Tremblay} et~al.}{2011}]{tremblay_11}
{Tremblay}, P.-E., {Bergeron}, P., and {Gianninas}, A. (2011) ,
\newblock {\em \apj} {\bf 730}, 128

\bibitem[\protect\astroncite{{Tremblay} et~al.}{2010}]{tremblay_10}
{Tremblay}, P.-E., {Bergeron}, P., {Kalirai}, J.~S., and {Gianninas}, A. (2010)
  ,
\newblock {\em \apj} {\bf 712}, 1345

\bibitem[\protect\astroncite{{Valenti} and
  {Fischer}}{2005}]{valenti_fischer_05}
{Valenti}, J.~A. and {Fischer}, D.~A. (2005) ,
\newblock {\em \apjs} {\bf 159}, 141

\bibitem[\protect\astroncite{{Valenti} and {Piskunov}}{1996}]{valenti_96}
{Valenti}, J.~A. and {Piskunov}, N. (1996) ,
\newblock {\em \aaps} {\bf 118}, 595

\bibitem[\protect\astroncite{{van Leeuwen}}{2007}]{van_leeuwen_07}
{van Leeuwen}, F. (2007) ,
\newblock {\em \aap} {\bf 474}, 653

\bibitem[\protect\astroncite{{Vogt}}{1987}]{vogt_87}
{Vogt}, S.~S. (1987) ,
\newblock {\em \pasp} {\bf 99}, 1214

\bibitem[\protect\astroncite{{Vogt} et~al.}{1994}]{vogt_94}
{Vogt}, S.~S., {Allen}, S.~L., {Bigelow}, B.~C., {Bresee}, L., {Brown}, B.,
  {Cantrall}, T., {Conrad}, A., {Couture}, M., {Delaney}, C., {Epps}, H.~W.,
  {Hilyard}, D., {Hilyard}, D.~F., {Horn}, E., {Jern}, N., {Kanto}, D.,
  {Keane}, M.~J., {Kibrick}, R.~I., {Lewis}, J.~W., {Osborne}, J.,
  {Pardeilhan}, G.~H., {Pfister}, T., {Ricketts}, T., {Robinson}, L.~B.,
  {Stover}, R.~J., {Tucker}, D., {Ward}, J., and {Wei}, M.~Z. (1994) ,
\newblock In {\em Society of Photo-Optical Instrumentation Engineers (SPIE)
  Conference Series}.  ({D.~L.~Crawford \& E.~R.~Craine} ed.), Vol. 2198 of
  {\em Society of Photo-Optical Instrumentation Engineers (SPIE) Conference
  Series}, pp. 362--+

\bibitem[\protect\astroncite{{Williams} et~al.}{2004}]{williams_04}
{Williams}, K.~A., {Bolte}, M., and {Koester}, D. (2004) ,
\newblock {\em \apjl} {\bf 615}, L49

\bibitem[\protect\astroncite{{Williams} et~al.}{2009}]{williams_09}
{Williams}, K.~A., {Bolte}, M., and {Koester}, D. (2009) ,
\newblock {\em \apj} {\bf 693}, 355

\bibitem[\protect\astroncite{{Wizinowich} et~al.}{2000}]{wizinowich_00}
{Wizinowich}, P., {Acton}, D.~S., {Shelton}, C., {Stomski}, P., {Gathright},
  J., {Ho}, K., {Lupton}, W., {Tsubota}, K., {Lai}, O., {Max}, C., {Brase}, J.,
  {An}, J., {Avicola}, K., {Olivier}, S., {Gavel}, D., {Macintosh}, B., {Ghez},
  A., and {Larkin}, J. (2000) ,
\newblock {\em \pasp} {\bf 112}, 315

\bibitem[\protect\astroncite{{Wright}}{2005}]{wright_05}
{Wright}, J.~T. (2005) ,
\newblock {\em \pasp} {\bf 117}, 657

\bibitem[\protect\astroncite{{Zurlo} et~al.}{2013}]{zurlo_13}
{Zurlo}, A., {Vigan}, A., {Hagelberg}, J., {Desidera}, S., {Chauvin}, G.,
  {Almenara}, J.~M., {Biazzo}, K., {Bonnefoy}, M., {Carson}, J.~C., {Covino},
  E., {Delorme}, P., {D'Orazi}, V., {Gratton}, R., {Mesa}, D., {Messina}, S.,
  {Moutou}, C., {Segransan}, D., {Turatto}, M., {Udry}, S., and {Wildi}, F.
  (2013) ,
\newblock {\em ArXiv e-prints}

\end{thebibliography}
\end{small}

\end{document}